\documentclass[conference]{IEEEtran}
\IEEEoverridecommandlockouts
\usepackage{cite}
\usepackage{amsmath, amssymb, amsfonts}
\usepackage{tipa}
\usepackage{algorithm}
\usepackage{algpseudocode}
\usepackage{graphicx}
\usepackage{xcolor}
\usepackage{dblfloatfix}
\usepackage{textcomp}
\usepackage{xcolor}
\usepackage{authblk}
\usepackage{hyperref}
\usepackage{orcidlink}

\def\BibTeX{{\rm B\kern-.05em{\sc i\kern-.025em b}\kern-.08em
    T\kern-.1667em\lower.7ex\hbox{E}\kern-.125emX}}
\begin{document}

\title{The OCON model: an old but gold solution for distributable supervised classification\\
\thanks{This work was partially supported by the EU under the Italian NRRP of NextGenerationEU, partnership on “Telecommunications of the Future” (PE00000001 - program “RESTART”) and by OPTIMIST project funded from the EXCELLENT SCIENCE - Marie Skłodowska-Curie Actions under grant agreement n° 872866}
}

\author[1]{Stefano Giacomelli \orcidlink{0009-0009-0438-1748}}
\author[1]{Marco Giordano \orcidlink{0009-0001-1649-6085}}
\author[2]{Claudia Rinaldi \orcidlink{0000-0002-1356-8151}}

\affil[1]{DISIM - University of L'Aquila, L'Aquila, Italy}
\affil[2]{CNIT - Consorzio Nazionale Interuniversitario per le Telecomunicazioni, L'Aquila, Italy}

\maketitle
\thispagestyle{plain}
\pagestyle{plain}
\begin{abstract}
This paper introduces to a structured application of the \textit{One-Class} approach and the \textit{One-Class-One-Network} model for supervised classification tasks, specifically addressing a \textit{vowel phonemes classification} case study within the Automatic Speech Recognition research field. Through \textit{pseudo}-Neural Architecture Search and Hyper-Parameters Tuning experiments conducted with an \textit{informed} grid-search methodology, we achieve classification accuracy comparable to nowadays complex architectures (90.0 - 93.7\%). Despite its simplicity, our model prioritizes generalization of language context and distributed applicability, supported by relevant statistical and performance metrics. The experiments code is openly available at our \href{https://github.com/StefanoGiacomelli/Vowel\_phonemes\_Analysis\_and\_Classification\_by\_means\_of\_OCON\_rectifiers\_Deep\_Learning\_Architectures/tree/main}{GitHub}.
\end{abstract}


\section{Introduction}
The intricate tapestry of human speech is woven through the delicate interplay of various phonemes, each contributing its unique acoustic signature to the rich spectrum of linguistic expression. Among these, vowels stand as fundamental building blocks, playing a pivotal role in shaping the intelligibility and emotive nuances of language. The classification of vowels within the phonetic landscape has been a subject of enduring interest and significance in the realm of linguistics and speech science, due to the broad range of applications, e.g.: language learning and pronunciation assessment \cite{asif2022}, dialectology and sociology \cite{liu2021}, forensic speaker identification \cite{amin2014}, assistive technologies \cite{Chandrakala2021}, emotion recognition \cite{caballero2013}, up to brain computer interface \cite{bci2021}.
Advancements in the field of vowel recognition have reached a \textit{state-of-the-art} (SoA), leveraging cutting-edge Machine Learning (ML) algorithms and sophisticated signal processing techniques to achieve unprecedented accuracy in deciphering and understanding vocal nuances. Techniques exploited for this purpose span from features extraction in time and transformed domains \cite{hasan2021}, to advanced ML solutions, from Neural Networks (NNs) to Deep Learning (DL) algorithms and models applied for general purpose Automatic Speech Recognition (ASR), as in \cite{10.1007/978-981-19-0604-6_17, 10.1007/s11042-023-16438-y}.

Our proposal aims to assess the reliability of a simplified Neural Architecture Search (NAS) and Hyper Parameters Tuning (HPs-T) combination for designing NNs feature abstraction layers. We propose a modular model, the ``One-Class One-Network'' (OCON), comprising parallelized binary classifiers focused on simpler phonetic recognition sub-tasks. We  evaluate available data constraints and task complexities w.r.t. the current SoA, aiming for a \textit{shallow} and optimizable architecture with a sustainable and straightforward (re)-training cycle. Additionally, we determine the minimum number of formant features needed to achieve current accuracy levels in phonetic recognition.

% --------------------------------------------------------------------------------------------------------------------------------
\section{Methodologies}
Following a standard DL experiment routine, we begin by gathering a reliable audio \textit{dataset} with heterogeneous phonetic representations and gender diversity among speakers: we initially constrained our linguistic research context to the \textit{General American English} case, as defined by the International Phonetic Association (IPA). Within this contextual dataset production, we encounter several good examples of pre-processed datasets (Sec. \ref{datasets}) where pre-arranged phraseological (or even specific words like \textit{/hVd/}, vowels between an "\textit{h}" and a "\textit{d}") speech segments were already recorded, analyzed and processed so as to elicit specific meaningful features: \textit{formant} frequencies. 

These features set represent the input data suitable for our NN model, and are usually retrieved by means of standardized key-steps, which can be summarized in: (\emph{1}) manually or automatically segment speech signals in \textit{semantic frames} according to a pre-arranged semantic grid (words/phonemes and silences); (\emph{2}) analyze isolated fragments by means of Linear Predictive Analysis/Coding (LPA/C) to extract a smoothed time-frequency spectral estimation (we are interested in the frame-by-frame aggregated spectral envelope); (\emph{3}) extract first $N$ spectral \textit{peaks} by means of whatever \textit{peaks estimation} algorithm keeping tracks (\textit{contouring}) of frame-wise continuities.

Next, we'll manipulate and refine these formant frequency tracks to generate a suitable data vector (\textit{features}) for the input stage of our NNs. This pre-processing step plays a vital role in enabling networks to learn abstract representations effectively, thereby maximizing recognition accuracy results.

\subsection{Datasets availability} 
\label{datasets}
% PB-database
From a preliminary dataset review, the initial \textit{/hVd/} vowel phoneme dataset, by Peterson and Burney (PB) \cite{10.1121/1.1906875} featured $10$ vowels spoken by $33$ men, $28$ women and $15$ children. While being crucial for foundational phonetics research e.g. \cite{10.1044/jshr.3604.694}, it faced criticism for its measurement and reporting limitations \cite{10.1121/1.398221}.

In contrast, the \textit{HGCW} database  \cite{10.1121/1.411872} includes $45$ men, $48$ women, $25$ boys, and $19$ girls' phoneme recordings, meticulously screened for non-native English, speech/language issues, respiratory infections, and inadequate hearing. Released openly in \texttt{.dat} format, it already includes formant analysis and fundamental frequency time contourings. Compared to the PB database, the HGCW dataset doubles the sample size, improves samples audio quality and addresses real-word complexity by including overlapping boundaries among available phoneme classes.

%TI-MIT
The Texas Instruments \& Massachusetts Institute of Technology (TI-MIT) \textit{Corpus of Read Speech} dataset \cite{10.35111/17gk-bn40}, originally developed for military research on telephone communications, is the most commonly used dataset for ASR applications. It comprises $6300$ utterances from $10$ sentences, spoken by $630$ speakers, with a notable gender imbalance ($70\%$ male, $30\%$ female). One of its most important spin-offs is the VTR database \cite{10.1109/ICASSP.2006.1660034}, which is specifically tailored for vocal tract resonances research and formant frequencies estimation. The core-test set of the VTR includes $192$ utterances, with $24$ speakers (each uttering $5$ phonemically-compact and $3$ phonetically-diverse sentences), while the training set consists of $346$ utterances with $173$ speakers (each uttering $1$ phonemically-compact and $1$ phonetically-diverse sentence).

Possible dataset references thus exist  for evaluating \textit{formant-based} phonetic classification algorithms, being often underutilized: researchers tend to favor occasional \textit{ad-hoc} sampling, hindering comparative and generalization studies. 

Recognizing limitations of existing datasets for robust generalized solutions, we opted for the HGCW dataset (which offers the higher phonetic complexity level) to expedite the data retrieval pipeline relying on pre-extracted formant data. In this way we aim to promote consistency with literature and streamline results evaluation.

\subsection{Features analysis \& processing}
\label{features_sec}
The HGCW dataset filename structure (Table \ref{hgcw_filenames}) contains encoded phonetic and speaker features crucial for preliminary statistical analysis.
\begin{table}
\begin{center}
\caption{HGCW dataset filenames structure}
\label{hgcw_filenames}
\begin{tabular}{| c | c | c | c |}
\hline
\textbf{$1^{st}$ \textbf{character}}& {$2^{nd} \& 3^{rd}$ \textbf{ch.s}} & {$4^{th} \& 5^{th}$ \textbf{ch.s}} & \textbf{Example}\\
\hline
m (\textit{man}) & spk. n° ($50$ tot.) & ARPABet ch.s & \texttt{m10ae}\\
b (\textit{boy}) & / ($29$ tot.) & / & \texttt{b11ei}\\
w (\textit{woman}) & / ($50$ tot.) & / & \texttt{w49ih}\\
g (\textit{girl}) & / ($21$ tot.) & / & \texttt{g20oo}\\
\hline 
\end{tabular}
\end{center}
\end{table}
Through ad-hoc \texttt{Python} scripts, we discovered that only $1597$ out of $1668$ samples could be effectively used for our task due to null features presence for some samples, caused by authors algorithm failures. To maintain learning consistency, these samples were filtered out, resulting in an additional under-representation of certain phoneme and speaker classes (Table \ref{hgcw_classes}). 
\begin{table}
\begin{center}
\caption{HGCW actual classes statistics}
\label{hgcw_classes}
\begin{tabular}{| c | c | c | c | c | c | c |}
\hline
\textbf{Phoneme} & \textbf{Samples} & \textbf{Boys} & \textbf{Girls} & \textbf{Men} & \textbf{Women} & \textbf{Label ID}\\
\hline
ae & $134$ & $25$ & $17$ & $45$ & $47$ & $0$\\
ah & $135$ & $24$ & $19$ & $45$ & $47$ & $1$\\
aw & $133$ & $24$ & $18$ & $45$ & $46$ & $2$\\
eh & $139$ & $27$ & $19$ & $45$ & $48$ & $3$\\
er & $118$ & $26$ & $18$ & $37$ & $37$ & $4$\\
ei & $126$ & $25$ & $17$ & $43$ & $41$ & $5$\\
ih & $139$ & $27$ & $19$ & $45$ & $48$ & $6$\\
iy & $124$ & $20$ & $18$ & $43$ & $43$ & $7$\\
oa & $136$ & $25$ & $19$ & $45$ & $47$ & $8$\\
oo & $139$ & $27$ & $19$ & $45$ & $48$ & $9$\\
uh & $138$ & $26$ & $19$ & $45$ & $48$ & $10$\\
uw & $136$ & $25$ & $19$ & $44$ & $48$ & $11$\\
\hline
\textbf{TOTAL} & $1597$ & $301$ & $221$ & $527$ & $548$ & $12$\\
\hline 
\end{tabular}
\end{center}
\end{table}
Fundamental frequency contours ($F0$) were estimated using a $2$-way auto-correlation/zero-crossing pitch tracker with a halving/doubling evaluation sub-routine \cite{10.1044/jshr.3604.694}, while formants were estimated using LPA spectra estimation and peak retrieval with parabolic interpolation \cite{10.2307/3680788}. The resulting frequency trajectories underwent further refinement using an interactive audio spectral editor for manual discontinuity examination and interpolation.

We created various experimental sub-structures of the original dataset, categorizing each sample based on:
(\emph{1}) \textit{Phonemes grouping}, including $F0$ and the first $3$ formant frequencies obtained at the \textit{steady state} (SS);
(\emph{2}) \textit{Speaker grouping}, segregating samples by gender (\textit{men}, \textit{women}, and \textit{children}) with the same features as before;
(\emph{3}) \textit{Phonemes grouping}, incorporating $F0$ and a total of $12$ formant frequency values, with the first $3$ formants sampled at $10\%$, $50\%$, SS, and $80\%$ of the total duration of the vowel nucleus.

% Features processing & Classification SoA
We analyzed and summarized classification algorithms exclusively evaluated on the PB and/or HGCW datasets features, to establish a consistent reference baseline for our work (Table \ref{statistical_ML}): Linear Discriminant Analysis (LDA) \cite{10.1121/1.393381} and Generalized Linear Regression Models (GLM) \cite{10.21437/ICSLP.1992-167} emerged as the prominent and most involved statistical-ML approaches, combined with innovative formant features processing as the \textit{3D-auditory target zones} framework, expressed by means of the logarithm of formant distances \cite{10.1121/1.397862}. Other studies \cite{10.1016/0167-6393(85)90040-8} adopted canonical auditory frequency transforms including: the Bark scale \cite{10.1121/1.385079, 10.1121/1.385780}, a \textit{technical} Mel scale approximation \cite{10.1515/9783110873429}, and a \textit{lin-to-log} frequency approximation \cite{Koening}.
\begin{table}
\begin{center}
\caption{Phonemes \& Speaker recognition w. PB dataset\\ (GM = \textit{geometric mean})}
\label{statistical_ML}
\begin{tabular}{| c | c | c | c | c |}
\hline
\textbf{Task} & \textbf{Data scale}& \textbf{Processing}& \textbf{ML}& \textbf{Accuracy}\\
\hline
Phoneme & Hz & \textit{Jacknife} & LDA & 81.8\%\\
& Log & None & GLM & 87.4\%\\
& / & -GM($F0$), $\cdot 0.333$& LDA & 86.3\%\\
& / & -($\bar{F1}, \bar{F2}, \bar{F3}$) & LDA & 89.5\%\\
& Bark & None & GLM & 86.2\%\\
& / & \textit{Jacknife} & LDA & 85.7\%\\
& / & -GM($F0$)& LDA & 85.3\%\\
& / & -($\bar{F1}, \bar{F2}, \bar{F3}$) & LDA & 88.3\%\\
& ERBs& None & GLM & 86.8\%\\
& / & -GM($F0$), $\cdot 0.5$& LDA & 87\%\\
& / & -($\bar{F1}, \bar{F2}, \bar{F3}$) & LDA & 88.8\%\\
\hline
Speaker & Hz & None & LDA & 89.6\%\\
& Bark & None & LDA & 88\%\\
& / & $\Delta F_n$ & LDA & 41.7\%\\
\hline
\end{tabular}
\end{center}
\end{table}

NNs research in phonetic recognition has predominantly centered on using LPA coefficients \cite{10.5120/7777-0862} or spectral/\textit{cepstral} derived features \cite{Kohonen}, often employing far more complex convolutional and/or recurrent stages. The only phonetic-OCON research retrieved \cite{10.21437/ICSLP.1998-40} reported improvements exclusively over TI-MIT data, combined with LPC features. Due to this scarcity of comparative literature, we set a target \textit{average accuracy} of $90\%$ aiming to improve results reported in \cite{10.1044/jshr.3604.694, 10.1121/1.411872}.

\begin{figure}[!t]
\centering
\includegraphics[width=3.2in]{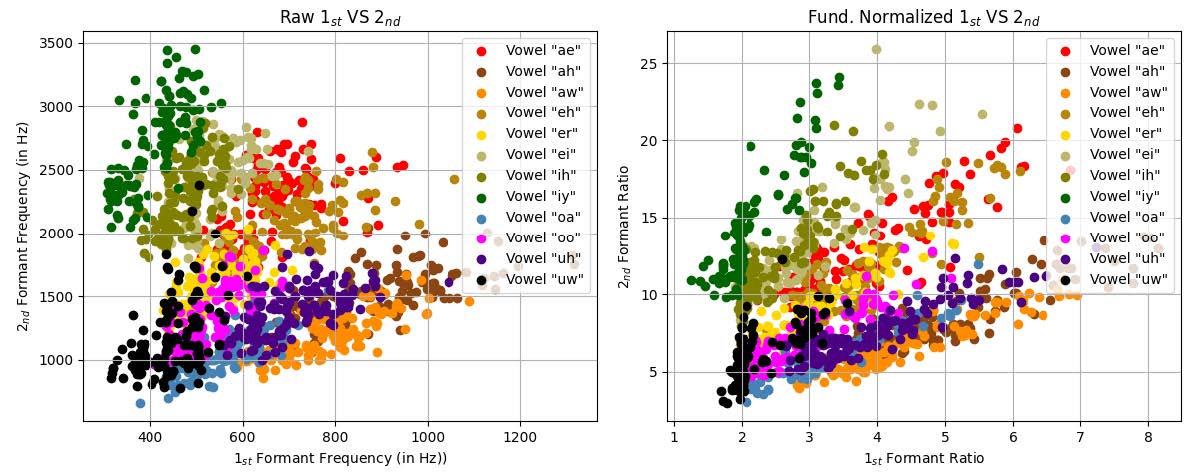}
\caption{HGCW dataset normalization ($2$D formantic projection)}
\label{hgcw_raw_vs_normalized}
\end{figure}

% Our features processing proposal
Considering the significant variation in $F0$s within speakers due to physiological factors and related pitch variations due to \textit{prosody}, we introduce linear formants normalization w.r.t. $F0$s. No prior usage of this pre-processing method were found, which appears to enhance class segregation by directly expressing distances in the linear frequency space (Fig. \ref{hgcw_raw_vs_normalized}).
To foster NNs training convergence, we applied \textit{min-max} scaling to normalize the entire feature set. We also assessed \textit{formant ratios} Probability Mass Distributions to determine Z-score (standardization) feasibility but they exhibited a consistent tendency towards skewed \textit{Poisson} or \textit{Log-normal} distributing. 

For enhanced data portability and reusability, we encoded all preprocessed features in a binary \texttt{NumPy} open-source compressed format (\texttt{.npz}) to preserve data resolution.

% --------------------------------------------------------------------------------------------------------------------------------
\section{Implementation}

In this section, we introduce the NN architectures for our experimentation, beginning with the Multi-Layer Perceptron (MLP). Since we are working with pre-processed features, our heuristic search experiments will focus on the architecture topology and characteristics of the MLP. %Nowadays, MLPs are used for feature abstraction in the classification stages of more complex models. 
We will then present our OCON proposal, which models multi-output classification tasks using multiple independent copies of the same optimized MLP architecture setup. These configurations are derived through simplified and \textit{informed} NAS (\textit{pseudo}-NAS) combined with HPs-T: in NNs research, HPs tuning involves optimizing architectural and learning parameters (such as layers, nodes, \textit{backpropagation} optimizers, learning rate etc.) to minimize the network cost function, between the predicted result (class) and the provided \textit{ground-truth} (label) in supervised learning contexts.

\subsection{Architecture \& Model}
\label{subsec:architecture_and_model}
MLPs, also known as \textit{fully connected} (FC)-layers or \textit{Feed-forward} NNs, are simply \textit{Perceptrons} (\textit{neurons}) stacked in vertical layers (\textit{shallow} NNs), whose function is: 
\begin{equation}
y_n=\varphi\langle x, w_k\rangle = x^\top w_k=\varphi\left(\sum_{k=0}^{K}x_nw_k\right)
\end{equation}
where $x_n$ are the input features, $w_k$ a set of scaling coefficients (\textit{weights}) and $\varphi(\cdot)$ a non-linear function (often \textit{activation}): in our case a standard ReLU \cite{10.1109/ICCV.2015.123}.

The One-Class-One-Network (OCON) model \cite{10.1109/ISCAS.1991.176636}, introduced in the '90s, served as a \textit{parallel distributed processing} solution to overcome limitations of architectures which required full re-training when altering dataset classes. Today, OCON resembles a simplified form of architecture \textit{ensembling} e.g. when multiple complex networks are combined through other networks or algorithms, to enhance overall models accuracy. In this case, a multi-output classification is distributed across independent sub-networks, each functioning as binary classifier: an approach widely used in anomaly detection and computer vision \cite{ DBLP:journals/corr/abs-1802-06360, 10.1109/TIP.2019.2917862}. In our case, we split a $12$-phoneme classification task into a bank of $12$ independent classifiers with identical architectural topology, seeking an \textit{optimal average} architecture estimate.

If a single output label is needed, a context-specific output algorithm must be devised. However, we find the classification \textit{logits} vector more beneficial for understanding phonetic class boundaries and feature complexities. 

While no literature references were found regarding OCON-specific output algorithms,
the \textit{argument of the maxima} (\textit{ArgMax}) approach can be employed, as it typically returns a single value representing the first occurrence of the maximum, when multiple occurrences exist.

\begin{figure}[!t]
\centering
\includegraphics[width=2.8in]{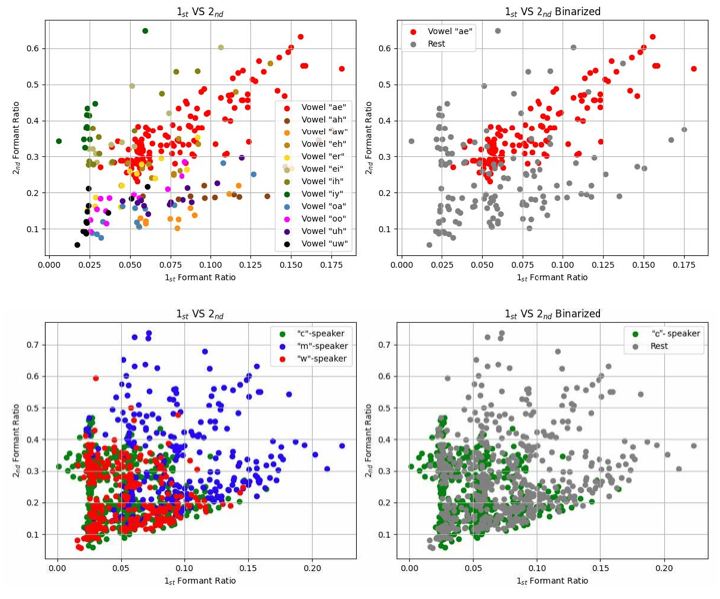}
\caption{HGCW dataset \textit{One-Hot encoding} examples}
\label{hgcw_one-hot}
\end{figure}

During supervised training, each sample label needs to be binarized (\textit{one-hot encoded}) according to the class architecture, while features are simultaneously fed to all classifiers. To accomplish this, we developed a custom one-hot encoding technique (Alg. \ref{one-hot_enc}) to transform labels based on the incoming \texttt{True}-One-class. Additionally, due to previous observations (Table \ref{hgcw_classes}), we needed to slightly down-sample resulting sub-sets, injecting further under-representation, to achieve nearly perfect training class balance.
\begin{algorithm}[H]
\scriptsize
\caption{HGCW One-Hot encoding}
\label{one-hot_enc}
\begin{algorithmic}
\Require $c$ \Comment{True-class index}
\Require $s$ \Comment{Phoneme groups size}
\Require $\mathcal{X}$ \Comment{Features dataset}
\Require $\mathcal{Y}$ \Comment{Dataset labels}
\State class$_1 =\mathcal{X}(c)$ \Comment{Initialize True-class subset}
\State size $=$ length$($class$_1)$ \Comment{Extract True-class size}
\State classes$_0 =$ list[ ] \Comment{Initialize False-class subsets}
\State sub-sizes $=$ $round(\frac{size}{11})$ \Comment{Compute False-classes size}
\For {$k$ in $\mathcal{Y}$} \do {} \Comment{Subsets selection loop}
    \If{$\mathcal{Y}_k \neq c$}
        \State class$_0 = rand(\mathcal{X}_k$, sub-sizes$)$ \Comment{Random downsampling}
        \State classes$_0$.append$($class$_0)$
    \Else
        \State pass
    \EndIf
\EndFor
\end{algorithmic}
\end{algorithm}
Alg. \ref{one-hot_enc} creates subsets of the HGCW dataset, with sizes determined by the \texttt{True}-class original size, ensuring balanced outliers by allocating the same number of samples for all \texttt{False}-classes, evenly distributed among the remaining $11$. If \texttt{True}-class sizes are not divisible by $11$, a variability of $1$ to $3$ samples was considered acceptable. The one-hot encoding routine is executed once per architecture training cycle, before \textit{train-eval-test} splitting and \textit{mini-batch} partitioning. Speaker-based encoding for \textit{male}, \textit{female}, and \textit{children} classes is achievable as well (Fig. \ref{hgcw_one-hot}).

\subsection{Pseudo NAS \& HPs-T heuristic search}
The term \textit{pseudo}-NAS refers to the \textit{a prìori} constraint applied to the architecture topology (MLP), as reported in Sec. \ref{subsec:architecture_and_model}. This evaluation assesses the optimal number of layers and nodes (per layer) required to effectively solve both phoneme and gender classification tasks. On the other hand, \textit{Grid-based} HPs search is a statistical method where all possible combinations of NNs HPs are independently sampled and evaluated through direct learning procedures. While theoretically effective, it can become time-consuming due to the exponentially increasing computational requirements for narrowing resolutions: in a standard procedure, all possible combinations must be tested before selecting the \textit{optimal} one. In our scenario, we achieved a good trade-off establishing independent resolutions for each HP beforehand, employing an \textit{informed} iterative approximation approach, outlined as follows:
(\emph{1}) define a specific HPs subset (not necessarily all at once, eventually fix others);
(\emph{2}) sample each HP with an arbitrary resolution;
(\emph{3}) test each HPs combination and resulting \textit{temporary best estimates} can be considered: \textit{inheritable} in following heuristic stages (\textit{optimal estimates}) or used to narrow parameters resolution sampling around \textit{local good estimates} (searching for better sets);
(\emph{4}) go to step (\emph{2}) and re-iterate as much as needed.

\noindent We acknowledge that this simplified approach \textit{roughly} approximates theoretical grid-search and that may lead to misleading local minima in the model costs. However our aim is to find an \textit{average} One-Class topology in a computationally feasible manner. Our heuristic learning experiments implies dataset partitioning into train ($70\%$), dev ($15\%$), and test ($15\%$) sets with seeded initial states. We measure accuracy and mini-batch training times, averaging results over a $3$-fold validation procedure for each One-Class
\begin{table}
\begin{center}
\caption{NAS \& Learning heuristic stage \\ IN = \textit{input nodes}, LR = \textit{learning rate}, HN/L = \textit{hidden nodes/layers}}
\label{experiment_1}
\begin{tabular}{| c | c | c |}
\hline
\textbf{Input Features} & \textbf{Fixed HPs}& \textbf{Testing HPs}\\
\hline
SS formant ratios & IN ($3$) & HN ($10, 50, 100$)\\
  & HL ($1$) & Backprop\\
  & activations (ReLU) & (\textit{Adam}, \textit{RMSProp})\\
  & states init. & LR\\
  & (\textit{standard} \cite{10.1109/ICCV.2015.123}, $b=0$) & ($10^{-3}, 10^{-4}, 10^{-5}$)\\
  & epochs ($1000$) & \\
  & batch size ($32$) & \\
  & k-folds ($3$)& \\
\hline
\textbf{TOT sets: 18} & \textbf{TOT architectures: 12} & \textbf{TOT cycles: 648}\\
\hline 
\end{tabular}
\end{center}
\end{table}

\begin{figure}[!t]
\centering
\includegraphics[width=3in]{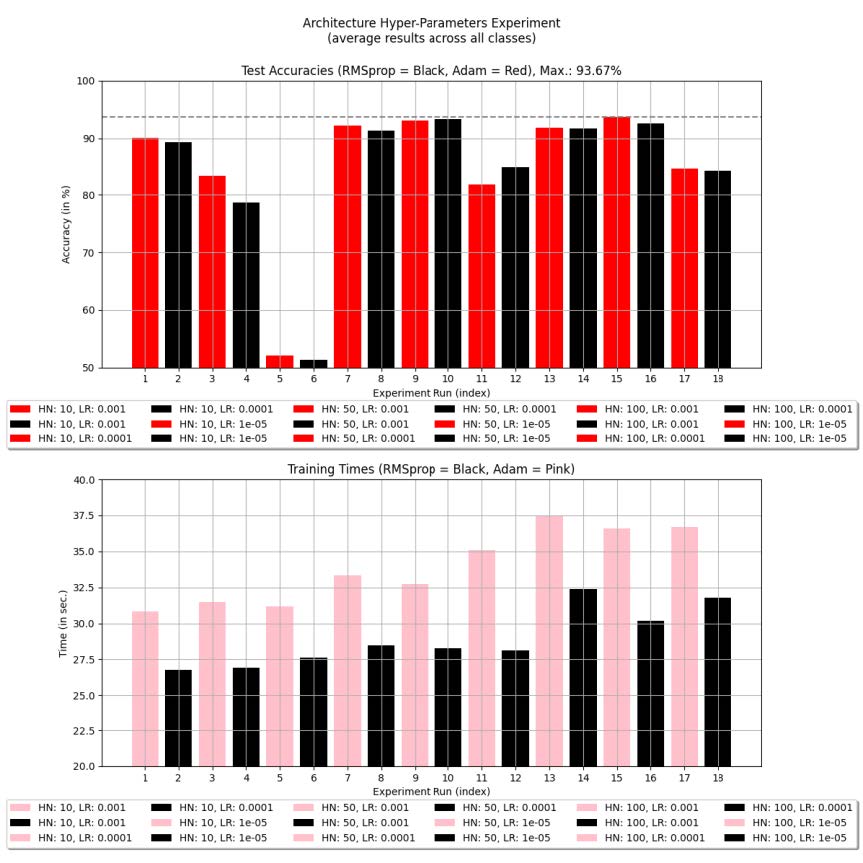}
\caption{$1st$ heuristic stage results}
\label{experiment_1_acc_times}
\end{figure}

In the $1^{st}$ heuristic stage (Table \ref{experiment_1}), which combined pseudo-NAS and HP-T experiments, two architectural combinations ($10^{th}$ and $15^{th}$) yielded similar average accuracies ($93.67\%$, Fig. \ref{experiment_1_acc_times}). The RMSProp optimizer \cite{Tieleman} appears to mitigate the increasing trend in learning times better than Adam \cite{Kingma} does, but we opted for the top-performing configuration (HL: $1$, HN=$100$, LR=$10^{-4}$, Backprop: Adam).

\begin{figure}[!t]
\centering
\includegraphics[width=3in]{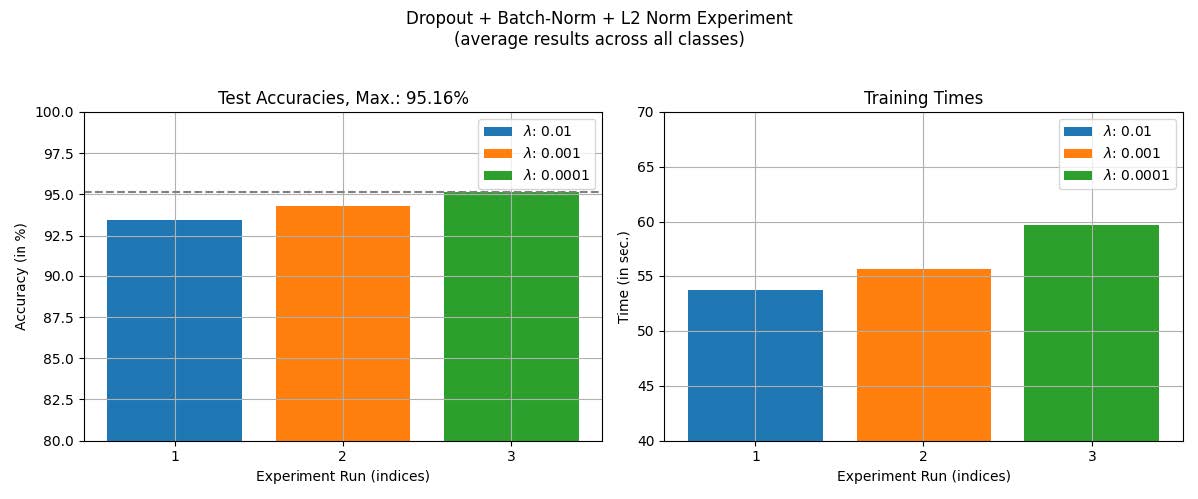}
\caption{$4^{th}$ heuristic stage results (L2-Norm)}
\label{experiment_4_acc_times}
\end{figure}

Next heuristic stages were structured to evaluate gradual introduction of \textit{regularization} techniques, evaluating potential benefits.
\begin{table}
\begin{center}
\caption{$2^{nd}, 3^{rd}$ \& $4^{th}$ heuristic stage (HP-T Regularization)\\ IN = \textit{input nodes}, LR = \textit{learning rate}, HN/L = \textit{hidden nodes/layers}}
\label{experiment_reg}
\begin{tabular}{| c | c | c |}
\hline
\textbf{DropOut HPs}& \textbf{Batch-norm HPs}& \textbf{L2-Norm HPs}\\
\hline
IN DropOut rate & LR & L2-Norm\\
(0.8, 0.9) & ($10^{-3}$, $10^{-4}$, $10^{-5}$) & $\lambda (10^{-2}, 10^{-3}, 10^{-4})$\\
HN DropOut rate & Batch-Norm & \\
($[0.5, 1.]$, res.: $0.1$) & & \\
\hline 
LR ($10^{-4}$) &  & LR ($10^{-4}$)\\
k-folds ($6$) & k-folds ($10$) & k-folds ($10$) \\
& batch size (32) & batch size (32) \\
epochs (3000) & epochs (1000) & epochs (1000) \\
\hline
\textbf{TOT cycles: 864} & \textbf{TOT cycles: 360} & \textbf{TOT cycles: 360}\\
\hline 
\end{tabular}
\end{center}
\end{table}
\begin{figure}[!t]
\centering
\includegraphics[width=3in]{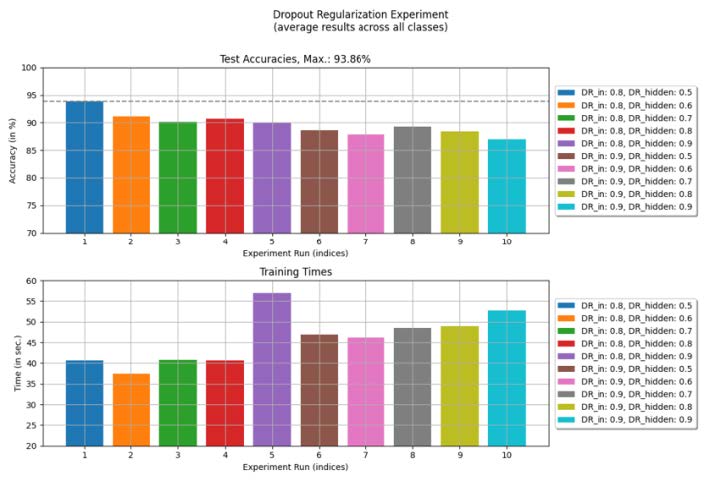}
\caption{$2^{nd}$ heuristic stage results (Dropout)}
\label{experiment_2_acc_times}
\end{figure}
In \textit{DropOut} \cite{10.5555/2627435.2670313} tests (Table \ref{experiment_reg}, Fig. \ref{experiment_2_acc_times}), we found that the best accuracy run (the $1^{st}$) is also among the fastest. We achieved $0.19\%$ increase in prediction accuracy at the expense of $+3.4$sec. in training time: resulting DropOut probabilities are $80\%$ for input nodes and $50$\% for hidden nodes.
\begin{figure}[!t]
\centering
\includegraphics[width=2.8in]{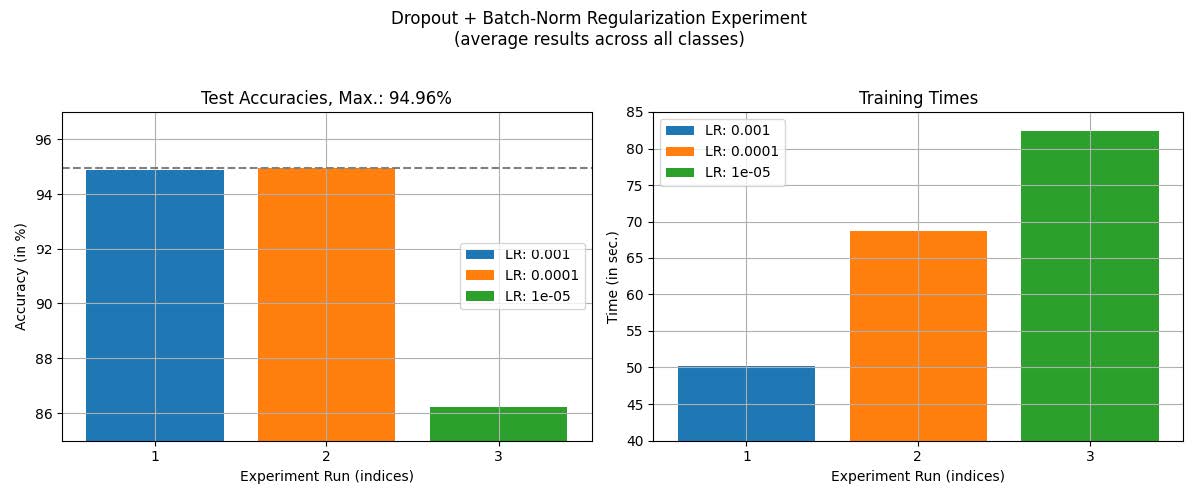}
\caption{$3^{rd}$ heuristic stage results (Batch-Norm)}
\label{experiment_3_acc_times}
\end{figure}
In \textit{Batch-norm} \cite{10.5555/3045118.3045167} tests, after re-evaluating LRs, it was confirmed that $10^{-4}$ yielded the best results: a significant $+1.1$\% in test accuracy, although average training times were nearly doubled. In \textit{L2-Norm} \cite{10.2307/1271436} (\textit{Ridge penalty}) tests (Table \ref{experiment_reg}), the best $\lambda$ (\textit{weight decay}) value was found to be $10^{-4}$ (Fig. \ref{experiment_4_acc_times}), resulting in a $+0.19\%$ in average accuracy. Interestingly, this improvement was accompanied by a decrease in training times (now below $60$sec.)
The resulting overall averaged One-Class proposal can be reviewed in Table \ref{one_class}.
\begin{table}
\begin{center}
\caption{One-Class architecture (MLP)\\ IL = \textit{input layer}, LR = \textit{learning rate}, HN/L = \textit{hidden nodes/layers},\\ ON = \textit{output nodes}}
\label{one_class}
\begin{tabular}{| c | c | c |}
\hline
\textbf{Architecture} & \textbf{Features} & \textbf{Learning}\\
\hline
IL: $3$ nodes & $\omega$ init.: \textit{Kaiming-He} norm. & \textit{Adam} optim.\\
HL: 1, $100$ nodes  & $b$ init.: $0$ & LR: $10^{-4}$\\
ON: 1 (\textit{logit}) & One-hot encoding & Mini-bacth\\
ReLU (common)& (Dataset \textit{re-shuffling}) & ($32$ samples)\\
\hline
IN-DR: $0.8$ & Batch-Norm  & L2-$\lambda$: $10^{-4}$\\
HL-DR: $0.5$ & & \\
\hline
\end{tabular}
\end{center}
\end{table}

% --------------------------------------------------------------------------------------------------------------------------------
\section{Model Training \& Results Discussion}
A parallelized set of independently trainable One-Class architectures was scripted and CPU runtimes (Google Colab) were utilized to efficiently measure isolated training cycle performance and resources consumption. The OCON architecture relies on the backprop loop of each MLP for consistent learning, while its \textit{inference} involves extracting sample features array, computing $12$ parallel one-hot encodings and conducting an ArgMax search to find the maximum value (predicted label) within the $12$-logit probabilities vector.

It follows a review of the \textit{phoneme recognition} experiments, where we tested the efficiency of each dataset sub-structure (Sec. \ref{features_sec})%: we remember that the One-Class estimate was built upon the simplest of the available structure ($3$xSS formant features, for each utterance)
. An \textit{Early-stopping} training strategy \cite{bai} was adopted, setting a $2$-variables-match escape condition: a \textit{minimum loss} threshold (averaging among last $50$ training samples loss) and a \textit{minimum test accuracy} threshold (w.r.t. last batch results). These variables were further empirically assessed to ensure practical convergence of training cycles, with each cycle not exceeding $25$-$30$min. While the learning phases may not be optimally exploited, they were deemed satisfactory for our study's purposes.

\subsection{Phonemes recognition}
Initially we re-evaluated the OCON model using the SS dataset variant (Table \ref{phonemes_1}):
\begin{table}
\begin{center}
\caption{$1^{st}$ Experiment: \textit{SS}-phonetic classification}
\label{phonemes_1}
\begin{tabular}{| c | c | c |}
\hline
\textbf{Features} & \textbf{Training} & \textbf{Early-Stopping}\\
\hline
SS formant ratios & epochs: $1000$& Loss thresh.: $0.2$\\
 & (for each \textit{batch-set}) & Accuracy thresh.: $90$\%\\
 & Re-shuffling & \\
 & balancing tol.: $0.01$ & \\
\hline
\textbf{Phonemes} & \textbf{Test Accuracy} (\%) & \textbf{Training times} (sec.)\\
\hline
ae & $86.27$ & $247.62$\\
ah & $90.85$ & $85.67$\\
aw & $86.09$ & $117.71$\\
eh & $89.05$ & $345.20$\\
er & $91.90$ & $25.71$\\
ei & $84.97$ & $539.79$\\
ih & $87.38$ & $207.92$\\
iy & $92.21$ & $33.78$\\
oa & $82.31$ & $120.20$\\
oo & $85.96$ & $396.59$\\
uh & $85.65$ & $485.34$\\
uw & $90.91$ & $219.23$\\
\hline
\textbf{AVG Acc.:} $87.79\%$ & \textbf{AVG Time:} $235.40$sec. & \textbf{OCON Acc.:} $70\%$\\
\hline
\end{tabular}
\end{center}
\end{table}

few loss functions and training accuracy curves reached visible \textit{plateaus}, with periodic artifacts (\textit{spikes}) indicating batch re-shuffling instances. %(Fig. \ref{spikes}) 
Surprisingly, the \textit{er} and \textit{iy} phoneme classes resulted well-represented, showing little or no changes in trend after encoding or re-shuffling: this suggests that under-represented classes remained the best represented (Table \ref{phonemes_1}). Classification accuracies were evaluated over the whole dataset (binary classification threshold set at $0.5$), with certain MLPs achieving a sufficient probabilities segregation: however, errors were high between \textit{aurally closest classes} (e.g.: \textit{ae} and \textit{eh}, \textit{er} and \textit{ei}).

Hidden dataset biases, such as \textit{children vs women utterances} (exceptional similarities in formantic disposition) were re-examined, leading to slight improvements in class boundaries upon filtering out (due to lack of insights about \textit{vocal maturity} of children samples) albeit increasing training duration: attempts to re-introduce $F0$s data to improve gender recognition were unsuccessful (AVG acc.: $88.80\%$, OCON acc.: $74\%$). The most performative features-set comprised temporal-tracks of formant ratios (Table \ref{phonemes_3}), leading to significant improvements in accuracy, reduced training times and mitigated Early-stopping side-effects, approaching the reference \cite{0.1088/1742-6596/2466/1/012008} accuracy goal of $90\%$ (Table \ref{roc_metrics}).
\begin{table}
\begin{center}
\caption{$3^{rd}$ Experiment: \textit{Time-Tracks}-phonetic classification}
\label{phonemes_3}
\begin{tabular}{| c | c | c |}
\hline
\textbf{Features} & \textbf{Training} & \textbf{Early-Stopping}\\
\hline
$10$\%, $50$\% & epochs: $1000$& Loss thresh.: $0.15$\\
SS, $80$\% & (for each \textit{batch-set}) & Accuracy thresh.: $95$\%\\
formant ratios & Re-shuffling & \\
 & balancing tol.: $0.01$ & \\
\hline
\textbf{Phonemes} & \textbf{Test Accuracy} (\%) & \textbf{Training times} (sec.)\\
\hline
ae & $94.55$ & $72.17$\\
ah & $91.80$ & $156.37$\\
aw & $89.86$ & $71.47$\\
eh & $93.74$ & $540.39$\\
er & $93.43$ & $28.05$\\
ei & $96.37$ & $104.98$\\
ih & $94.55$ & $97.20$\\
iy & $96.49$ & $38.59$\\
oa & $93.49$ & $49.43$\\
oo & $95.62$ & $96.17$\\
uh & $90.98$ & $649$\\
uw & $93.74$ & $108.30$\\
\hline
\textbf{AVG Acc.:} $93.72\%$ & \textbf{AVG Time:} $167.68$sec. & \textbf{OCON Acc.:} $90\%$\\
\hline
\end{tabular}
\end{center}
\end{table}

\begin{table}
\begin{center}
\caption{OCON Normalized ROC-AUC/DET metrics}
\label{roc_metrics}
\begin{tabular}{| c | c | c | c | c | c |}
\hline
\textbf{One-Class} & \textbf{ER} & \textbf{FDR} & \textbf{FOR} & \textbf{NPV} & \textbf{AUC}\\
\hline
ae & $0.02$ & $0.03$ & $0.01$ & $0.99$ & $0.9986$\\
ah & $0.03$ & $0.06$ & $0.00$ & $1.00$ & $0.9866$\\
aw & $0.03$ & $0.06$ & $0.00$ & $1.00$ & $0.9980$\\
eh & $0.02$ & $0.03$ & $0.01$ & $0.99$ & $0.9934$\\
er & $0.02$ & $0.03$ & $0.00$ & $1.00$ & $0.9935$\\
ei & $0.03$ & $0.05$ & $0.01$ & $0.99$ & $0.9979$\\
ih & $0.03$ & $0.05$ & $0.00$ & $1.00$ & $0.9996$\\
iy & $0.01$ & $0.02$ & $0.00$ & $1.00$ & $0.9994$\\
oa & $0.04$ & $0.07$ & $0.00$ & $1.00$ & $0.9898$\\
oo & $0.01$ & $0.01$ & $0.00$ & $1.00$ & $1.0000$\\
uh & $0.03$ & $0.05$ & $0.00$ & $1.00$ & $0.9950$\\
uw & $0.03$ & $0.06$ & $0.01$ & $0.99$ & $0.9965$\\
\hline
\end{tabular}
\end{center}
\end{table}

\section{Conclusions and future works}
We know that a single \textit{Perceptron} can actually predict a speech signal sample \cite{10.1109/ICNN.1995.487832} approximating LPA results: our model proposal could be considered then an \textit{ad-hoc} \textit{head}-integration for a complex \textit{Perceptron-based} formant neural framework. We are aware of researches on formants estimation \cite{ALKU2023101515, 10.1121/1.5088048} leveraging convolutional and recurrent layers (\textit{backbone stages}): despite accuracies achieved, we believe our approach, employing pseudo-NAS/HP-T techniques completely hand-scripted and fully conducted on Colab notebooks, could be broaderly re-applied to effectively assess newer SoA NNs building blocks efficiency, in terms of parameters reduction.

Our model exhibits high distributability, with each classifier independently re-trainable and sufficiently lightweight to be suitable for constrained computational contexts and integration into complex architectures. Optimization techniques, like \textit{parameters prunings} and \textit{quantization} could further enhance its memory consumption at inference time. Additionally, its modular structure facilitates adaptation into different language contexts. We challenge the notion that larger datasets or models inherently yield better accuracies, asserting that our approach offers good generalizability, despite observed limitations in training sample size. We encountered difficulty finding more extensive pre-processed datasets but we aim to validate our findings by expanding our dataset sources, potentially including datasets like TI-MIT, UCLAPhoneticsSet and/or AudioSet. Our proposal for linear features processing demonstrates that altering speech signal spectra in auditory-based non linear ways isn't always the optimal method for descriptive speech modeling. However, we intend to re-examine solutions from existing literature.

%Regarding sustainability, we're pleased to find that the CO2e emissions for fully retraining our model are just over half the emissions from the entire lifecycle of a single cigarette. 

Future research could focus on enhancing label selection, considering the potential for \textit{increased reliability} by applying \textit{training assurance} scaling coefficients to output One-Class probabilities: this could involve analyzing epochs spent by the classifier to maintain loss below the specified Early-stopping threshold and considering derivatives of the loss curve to further refine the output probabilities, especially in cases of too rapid training error minimization.

\section*{Acknowledgments}
\noindent We express our gratitude to the Electronic Music (meAQ) Dpt. at A. Casella Conservatory (L'Aquila) for their assistance in developing this project. ChatGPT (OpenAI) was used here to support typos correction and text formatting.

\bibliographystyle{IEEEtran}
\bibliography{References}

\end{document}